\documentclass[runningheads]{llncs}
\usepackage{graphicx}

\newenvironment{blockquote}{%
  \par%
  \medskip
  \leftskip=1.5em\rightskip=1.5em%
  \noindent\ignorespaces}{%
  \par\medskip}
  
\begin{document}
\title{A Comparative Study of Learning Outcomes for Online Learning Platforms}
%
%
\author{Francois St-Hilaire \inst{1} \and Nathan Burns \inst{1} \and Robert Belfer \inst{1} \and Muhammad Shayan \inst{1} \and Ariella Smofsky \inst{1} \and Dung Do Vu \inst{1} \and Antoine Frau \inst{1} \and Joseph Potochny \inst{1} \and Farid Faraji \inst{1} \and Vincent Pavero \inst{1} \and Neroli Ko \inst{1} \and Ansona Onyi Ching \inst{1} \and Sabina Elkins \inst{1} \and Anush Stepanyan \inst{1} \and Adela Matajova \inst{1} \and Laurent Charlin \inst{2} \and Yoshua Bengio \inst{2} \and Iulian Vlad Serban \inst{1} \and Ekaterina Kochmar \inst{3}}

%
%
\institute{Korbit Technologies Inc., Canada \and
Quebec Artificial Intelligence Institute (Mila), Canada \and
University of Bath, United Kingdom}
%
\maketitle              
\begin{abstract}

Personalization and active learning are key aspects to successful learning.
These aspects are important to address in intelligent educational applications, as they help systems to adapt and close the gap between students with varying abilities, which becomes increasingly important in the context of online and distance learning. We run a comparative head-to-head study of learning outcomes for two popular online learning platforms: {\tt Platform A}, which follows a traditional model delivering content over a series of lecture videos and multiple-choice quizzes, and {\tt Platform B}, which creates a personalized learning environment and provides problem-solving exercises and personalized feedback.
We report on the results of our study using pre- and post-assessment quizzes with participants taking courses on an introductory data science topic on two platforms.
We observe a statistically significant increase in the learning outcomes on {\tt Platform B}, highlighting the impact of well-designed and well-engineered technology supporting active learning and problem-based learning in online education. 
Moreover, the results of the self-assessment questionnaire, where participants reported on  perceived learning gains,  suggest that participants using {\tt Platform B} improve their metacognition.

\end{abstract}

\section{Introduction}

\begin{figure}[t!]
\centering
\includegraphics[width=0.8\textwidth]{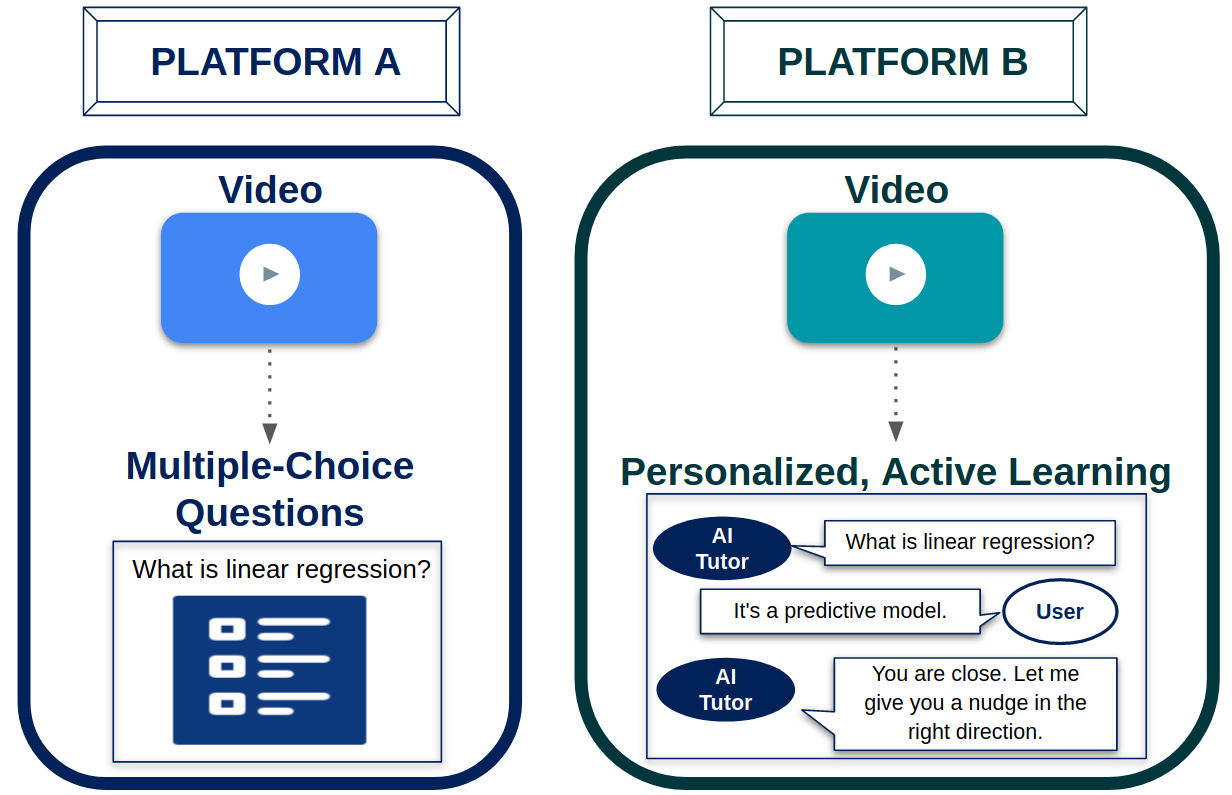}
\caption{{\tt Platform A} follows a traditional learning approach, while {\tt Platform B} uses a personalized, active learning approach with problem-solving exercises.} \label{fig:platforms}
\end{figure}

We investigate the learning outcomes induced by two popular online learning platforms in a comparative head-to-head study.
{\tt Platform A} is a widely used learning platform that follows a traditional model for online courses: students on this platform learn by watching lecture videos, reading, and testing their knowledge with multiple choice quizzes.
In contrast, {\tt Platform B}\footnote{{\tt Platform B} is the Korbit learning platform available at \url{www.korbit.ai}.} 
takes a different approach to online learning, focusing more on active learning and personalization for the students~\cite{Serban}.
{\tt Platform B} is powered by an AI tutor, which creates a personalized curriculum for every student and teaches through short lecture videos, interactive problem-solving exercises, mini-projects, and by providing personalized pedagogical interventions mimicking a human tutor. Specifically, {\tt Platform B} alternates between lecture videos and interactive problem-solving exercises, during which the AI tutor poses questions and shows students problem statements, and students attempt to solve the exercises. Students can also pose their own questions on the material, ask for help, or even skip exercises.
The AI tutor addresses each incorrect attempt and each request for help with one of a dozen pedagogical interventions, which are tailored to student's needs, thus ensuring that interactions are personalized. Figure \ref{fig:platforms} visualizes the differences between the two platforms.

The goal of this study is to 
measure the efficiency with which students learn on each platform.
Since the key difference between the platforms is 
the fact that {\tt Platform B} supports personalized, active learning and problem-based learning, our aim is to investigate to what extent such mode of tutoring on online platforms contributes to learning outcomes.
We aim to test the following hypothesis:

\begin{blockquote}
{\bf Hypothesis}: 
Participants who take the course on {\tt Platform B} have higher learning gains than those who take the course on {\tt Platform A}, because {\tt Platform B} provides a wider and more personalized variety of pedagogical elements to its students.
\end{blockquote}

\section{Related Work}


Online learning platforms providing a massive number of students with access to learning on various subjects have the potential to revolutionize education~\cite{Graesser,Koedinger,Psotka,Wang}. In particular, such platforms have the capability of bridging the gap and addressing inequalities in the society caused by uneven access to in-person teaching~\cite{Hrastinski,Tomkins}. The current pandemic only exacerbates the need for the high quality online education being accessible to a wide variety of students~\cite{Adedoyin,Basilaia,Onyema}.


Nevertheless, the efficacy of online and distance learning has been and continues to be challenged by researchers.
It was found that the course design and the mode of teaching strongly influence the way in which students progress~\cite{Tomkins,Vigentini}.
Specifically, it may be hard to address the differences in students' learning needs, styles and aptitudes~\cite{Coffield,Honey,Stash,VanLehn2007}, and this calls for approaches that can be adapted and personalized to the needs of each particular student.
Studies confirm that personalization is key to successful online and distance learning~\cite{Narciss,Sampson}, such as personalized complexity level and personalized feedback, as it can maximize the learning benefits for each individual student~\cite{Yin}. 

A number of studies have demonstrated that problem-solving is a highly effective approach for learning in various domains~\cite{Fossati,Kolb,Kumar_2005a,Wood,Woolf}.
Such problem-solving learning activities can be addressed by intelligent tutoring systems, which are also capable of giving personalized feedback and explanations, incorporating conversational scaffolding, and engaging students into active and problem-solving exercises~\cite{Albacete,budenbender2002,Chi,Fossati,Kumar_2005b,Lin,melis2004,Munshi,Nye,Rus_14a,Rus_14b}.

Many studies have been conducted evaluating the impact of educational technology and online learning platforms on student learning outcomes \cite{Demmans,Kashihara,Ma2014,Mark1993} \cite{Penstein2001,Tan,VanLehn2007,VanLehn2011}.
We adopt the well-established pre-post assessment framework, where students are split into intervention groups and their knowledge of the subject is evaluated before and after their assigned intervention.

In contrast to previous studies investigating learning outcomes with intelligent tutoring systems, in this study the AI-powered learning platform, {\tt Platform B}, is a fully-automated system based on machine learning models~\cite{Serban}. The system is trained from scratch on educational content to generate automated, personalized feedback for students and has the ability to automatically generalize to new subjects and improve as it interacts with new students~\cite{Kochmar,Grenander}.



In the context of online and distance learning, students' ability to self-assess, develop self-regulation skills and strategies, plays a crucial role~\cite{Barokas,Kashihara,Munshi}. However, many studies show that students generally struggle to evaluate their own knowledge and skills level~\cite{Brown_13,Brown_15,Crowell}.



\section{Experimental Setup}
\label{sec:setup}

\subsubsection{Participants}

$48$ participants completed a $3$-hour long course on {\em linear regression} using one of the two online platforms.
Their learning outcomes were measured before and after the course using pre- and post-assessment quizzes.
The experiment was run completely online. 
Participants completing either course were rewarded a $\$200$ Amazon gift card.


To recruit participants, we posted ads on social media and sent out emails to student clubs from local universities. Candidates interested in participating had to fill out a questionnaire specifying their field of study, their degree, and whether they have completed courses in machine learning or artificial intelligence.



Candidates were classified as eligible or ineligible based on their answers to this enrollment questionnaire. Specifically, candidates who had or were studying a math-heavy discipline at university (e.g. mathematics, statistics, physics) or who had completed any courses on statistics, machine learning or artificial intelligence were deemed ineligible.
As a result, out of the $60$ applicants $48$ participants were selected. The majority fall into our target audience of undergraduates ($89.6\%$) studying disciplines not centered around mathematics: health sciences ($27.7\%$), computer science ($23.4\%$), cognitive science ($12.8\%)$, among others. 




To ensure unbiased setting for the experiments on two platforms, we randomly divided participants into two groups.
The first group was asked to study the course on linear regression from {\tt Platform A}, and the second was asked to study the course on the same subject from {\tt Platform B}. 

\subsubsection{Choice of the Material}

Linear regression was selected as the topic of study on both online platforms since it is one of the most fundamental topics, that is covered early on in any course on machine learning and data science, and the material covering this topic on both platforms is comparable. To yield a fair comparison between the two platforms, extra care was taken to ensure that the linear regression courses were as similar as possible and that the sub-topics covered, the difficulty level, and the length of both courses were carefully aligned. 

The linear regression course on {\tt Platform B} was adapted for this study, combining existing and new content specifically created to align with the sub-topics covered in the {\tt Platform A} course. As a result, the courses on both platforms contain an introductory session and provide short lecture videos, followed by multiple-choice questions in the case of {\tt Platform A} and interactive problem-solving exercises in the case of {\tt Platform B}. The sub-topics taught include numerical variables, correlation, residuals, least squares regression, and evaluation metrics. 
The course on each platform takes approximately 3 hours to complete.



\subsubsection{Study Flow}


The study ran over a $4$-day period with strict deadlines set for the participants. The participants received instructions detailing all the steps they would need to complete on day $1$, and from this point they had $3$ days to complete the course. If they completed all necessary steps before the day $3$ deadline, they were asked to complete the final post-quiz by the end of day $4$. 

All participants were required to take an assessment quiz on linear regression before the course ({\em pre-quiz}) and another one after the course ({\em post-quiz}). Using pre- and post-quiz scores, we measure {\em learning gains} to quantify how efficiently each participant has learned. The pre- and the post-quizzes both consisted of $20$ multiple-choice questions and were equally adapted to both courses, meaning that any topic or concept mentioned in the quizzes was covered in both courses to an equal extent, ensuring that students mastering the topics using either course would be able to succeed in answering them.
In addition, the quizzes went through an independent review process, which ensured that the quiz scores were not inherently biased towards one of the learning platforms. Furthermore, each question of the pre-quiz was isomorphically paired with a question in the post-quiz, meaning that the difficulty of the two quizzes was as similar as possible without any questions being identical. This ensured that learning gains were accurately measured without any bias from the differing quiz difficulty.\footnote{Full content of the assessment quizzes will be made public upon paper acceptance.}

%

\subsubsection{Safeguarding Against Invalid Results}

Since the study ran fully online, it was important to take precautionary measures to minimize the chances of the participants cheating or otherwise not following the instructions. 
We identified the following two potential scenarios: first, a participant might have completed the pre- and the post-quizzes without really studying the course, or going through the course without paying attention, skipping the videos and exercises; second, a participant might have used external resources to find the correct answers on the quizzes. In both cases the participant's scores would be meaningless because it would be completely or almost completely unrelated to the course and the learning platform. To minimize the chances of this happening, we provided participants with very clear instructions on what they were required and what they were not allowed to do, and we required them to upload a completion certificate confirming that they at least went through the whole course on the corresponding platform. 
We also expect that scheduling the study over $4$ days, regardless of how quickly the participants could go through the course, helped discourage them from breezing through the quizzes and the course to immediately get their gift card and ensured there was a delay between their learning and assessment.
For {\tt Platform B} involving problem-solving exercises, we further defined a minimum requirement for participants to attempt at least $80\%$ of the exercises.

\subsubsection{Learning Gains}
To evaluate which of the two online learning platforms teaches the participants more effectively, we compare {\tt Platform A} and {\tt Platform B} based on the {\em average learning gain} and {\em normalized learning gain}~\cite{Hake} of the participants on each platform. A student's learning gain $g$ is estimated as the difference between their score on the post-quiz and on the pre-quiz as follows:

\begin{equation}
g = post\_score - pre\_score
\end{equation}

\noindent where $post\_score$ is the student's score on the post-quiz, and $pre\_score$ is their score on the pre-quiz. Both scores fall in the interval $[0\%, 100\%]$. A student’s individual normalized learning gain $g_{norm}$ is calculated by offsetting a particular student's learning gain against the score range in the ideal scenario in which a student achieves a score of $100\%$ in the post-quiz:

\begin{equation}
g_{norm} = \frac{post\_score - pre\_score}{100\% - pre\_score}
\end{equation}

\section{Results and Discussion}

$25$ participants completed the course on {\tt Platform A} and $23$ participants completed the course on {\tt Platform B}.
One participant on {\tt Platform B} did not satisfy the requirement of attempting at least $80\%$ of the exercises and was therefore excluded from the analysis.



It is worth noting that in order to align the courses on the two platforms closely, the personalized curriculum and the programming exercises offered by {\tt Platform B} were specifically disabled for this study. On the one hand, this allowed us to compare the two platforms on a fair basis and specifically explore the effects of personalized, active learning and problem-based exercises on the learning outcomes in accordance with the formulated hypothesis. On the other hand, given that the personalized curriculum and programming exercises are highlighted on the {\tt Platform B} website, this may have created a mismatch between students' expectations and their actual learning experience with this platform. Future experiments using the full functionality of {\tt Platform B} will aim to investigate the effect of other pedagogical elements.

\begin{figure}[ht!]
\includegraphics[width=\textwidth]{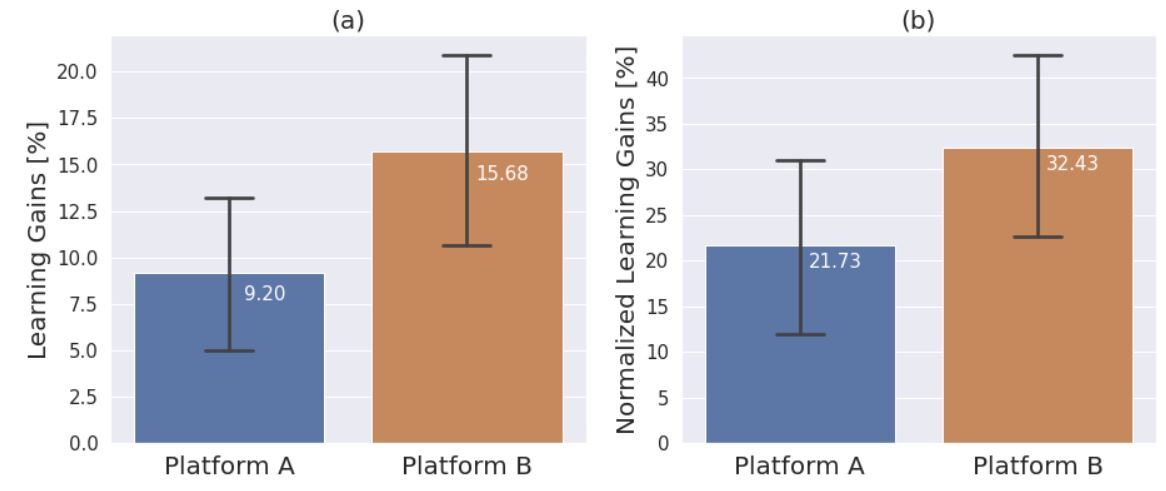}
\caption{(a) Average learning gains $g$ with $95\%$ confidence intervals.$^*$ (b) Average normalized learning gains $g_{norm}$ with $95\%$ confidence intervals.$^{**}$
Here $^*$ and $^{**}$ indicate a statistically significant difference at $95\%$ and $90\%$ confidence level respectively.} \label{fig:results}
\end{figure}

\subsubsection{Learning Outcomes}

Average learning gains are shown in Figure \ref{fig:results} for the two learning platforms.\footnote{$95\%$ confidence intervals (C.I.) are estimated as: $1.96 * \frac{{Standard \ Deviation}}{\sqrt{{Population \ Size}}}$.}
The average normalized learning gains $g_{norm}$ for {\tt Platform B} participants are $49.24\%$ higher than the average normalized gains for {\tt Platform A} participants. This difference is statistically significant at the $90\%$ confidence level ($p$=$0.068$). When considering raw learning gains $g$, the {\tt Platform B} average is $70.43\%$ higher than the {\tt Platform A} average with $95\%$ confidence ($p$=$0.038$).

It should be observed that there are $3$ participants from {\tt Platform A} and 2 from {\tt Platform B}, who showed negative learning gains.
These results are hard to interpret because the difficulty level of the pre- and the post-assessment quizzes is the same.
A possible explanation for this is that they had little prior knowledge and did not learn the material during the course, but managed to pick the correct answers in the pre-quiz by chance and were less lucky in the post-quiz.

We also observe that the median normalized learning gain $g_{norm}$ is only slightly higher for {\tt Platform B}, being at $31.82\%$ as opposed to $28.57\%$ for {\tt Platform A}. With the {\tt Platform B} average being substantially higher, this could mean that participants who underperform on {\tt Platform A} are well below the {\tt Platform A} median, and that participants who overperform on {\tt Platform B} are well above the {\tt Platform B} median. 
This might stem from the fact that {\tt Platform B} is better adapted to students with different backgrounds and learning needs by having various pedagogical tools and more active learning. 
However, we believe that more data and further studies would be needed to confirm this assumption. 

Further, the medians for raw learning gains $g$ are closer to the respective average values for both platforms: $10\%$ for {\tt Platform A} and $15\%$ for {\tt Platform B}. This is in line with our hypothesis, but suggests that while the normalization process produces a more meaningful metric for learning gains, it does introduce a lot of variance in the metric due to dividing by the pre-quiz score.

Overall, our hypothesis that learning outcomes are higher for participants on {\tt Platform B} than participants on {\tt Platform A} is confirmed by the results presented here, with the average learning gains $g$ and normalized learning gains $g_{norm}$ being substantially higher for {\tt Platform B} at $95\%$ and $90\%$ statistical significance-level respectively. To explain these results, we theorize that the active learning elements of {\tt Platform B} play a significant role in the participants' learning experience. The key difference between the two platforms is that {\tt Platform B} supports active learning and problem-based learning, while {\tt Platform A} is limited in this regard. To give an idea of the extent to which {\tt Platform B} provides more active learning than {\tt Platform A}, we estimated how much time participants spent on different learning activities while completing the course on either platform. Table \ref{tab:active_time} presents the estimates. 


\begin{table}
\centering
\caption{Estimated time spent on different learning activities on each platform (minutes and \% of total), excluding time spent on enrollment and assessments}\label{tab:active_time}
\begin{tabular}{|l|l|l|}
\hline
\textbf{Learning Activity} & \textbf{{\tt Platform A}} & \textbf{{\tt Platform B}} \\ \hline
Watching lecture videos & $46.0$ min. ($33.82\%$) & $29.6$ min. ($24.89\%$) \\ \hline
Reading material           & $50.0$ min. ($36.76\%$) & $0.0$ min. ($0.00\%$)     \\ \hline
Solving quizzes and exercises \ & $40.0$ min. ($29.42\%$) & $89.3$ min. ($75.11\%$) \\ \hline
\end{tabular}
\end{table}

For {\tt Platform A}, the time estimates for videos and readings come directly from the course website. For the exercises, {\tt Platform A} gives no time estimate for the 13 multiple choice questions in the course. We estimated it at 40 minutes by testing it ourselves. For {\tt Platform B}, we have access to the data from the participants' interactions on the platform, which lets us directly calculate estimates for the average time spent watching videos and solving exercises. Based on these estimates, we conclude that participants on {\tt Platform B} spent more time on active learning than {\tt Platform A} participants by a factor of $2.23$. We believe that this substantial difference in the active learning time is one of the main reasons for the higher learning gains observed w.r.t.\@ {\tt Platform B} participants. 

We further investigate the participants' behavior on {\tt Platform B} using the data collected automatically on this platform and report the most insightful correlations observed between participants' behavioral factors and their learning gains. Firstly, we establish a correlation between the total amount of time participants spent on solving exercises and their overall performance on exercises. Specifically, there is a positive correlation between the time spent on exercises and the rate at which participants provided correct answers on the first try ($r$$=$$0.34$). At the same time, the time spent on exercises and the average number of attempts participants needed to get a correct answer are negatively correlated ($r$$=$$-0.34$). This suggests that participants who took more time working on the exercises and formulating their answers performed better. These participants got the correct answer on the first try more often and, on average, took fewer attempts to answer correctly.
Secondly, we observe a correlation between the participants' performance on exercises and their learning gains.
Specifically, the rate of correct answers on the first try positively correlates with both learning gains ($r$$=$$0.44$) and post-quiz results ($r$$=$$0.46$), and the number of exercises completed positively correlates with the post-quiz score ($r$$=$$0.28$). These correlations may suggest that participants who spent more of their study time on active learning and problem-solving exercises performed better and, as a result, obtained higher post-quiz scores and learning gains.
We believe this might indicate the effectiveness of active learning elements on {\tt Platform B}.
One limitation to this hypothesis is that no strong correlation was found between the time spent on exercises and the learning gains or post-quiz scores. We believe this is partly due to the diversity in the background knowledge of the participants, leading to high variance in the pace at which participants completed exercises. More data is needed to thoroughly understand the factors driving the learning gains.

Another interesting observation is that participants who scored higher on the pre-quiz were less receptive to the pedagogical interventions from the AI tutor on {\tt Platform B}. Specifically, for participants who scored over $50\%$ on the pre-quiz, there is a stronger negative correlation between their learning gains and the total number of hints they received during the course ($r$$=$$-0.58$). For participants who scored below $50\%$ on the pre-quiz, this correlation is weaker ($r$$=$$-0.32$).
This suggests that the pedagogical interventions were more helpful for the students with lower pre-quiz scores.
We believe that the negative correlation between the number of hints received and learning gains is partly explained by the fact that pedagogical interventions on {\tt Platform B} mostly occur when a student is already struggling with an exercise. Therefore participants who received more hints most likely struggled more with the exercises, which, as discussed above, correlates with lower learning gains. 

\subsubsection{Metacognitive Evaluation}

In addition to measuring actual learning gains using pre- and post-quizzes, we evaluated various metacognition aspects related to the students' learning experience with the two platforms using a questionnaire where students were asked to report on their experience. Table \ref{tab:perceived} lists $3$ questions we asked the participants on their {\em perceived learning gains}. 

\begin{table}
\centering
\caption{Average perceived learning gains ($\pm$ $95\%$ confidence intervals).}\label{tab:perceived}
\begin{tabular}{|l|l|c|c|}
\hline
{\bf ID} & {\bf Question} & {\tt Platform A} & {\tt Platform B}\\
\hline
Q1 & How would you rate your comprehension & $4.16 \pm 0.27$ & $3.65 \pm 0.29 $ \\
& of the topics you studied? ($1$-$5$) & & \\ \hline
Q2 & How well do you think you performed & $76.7 \pm 5.41\%$ & $75.2 \pm 5.63\%$\\
& on the final quiz? ($1$-$100\%$) & & \\ \hline
Q3 & How capable would you feel in applying & $3.68 \pm 0.29$ & $3.35 \pm 0.38$ \\
& the skills you learned in a practical setting & & \\
& (at your job, in a personal project, in your & & \\
& research, etc.)? ($1$-$5$) & & \\
\hline
\end{tabular}
\end{table}

\begin{figure}[ht!]
\centering
\includegraphics[width=0.8\textwidth]{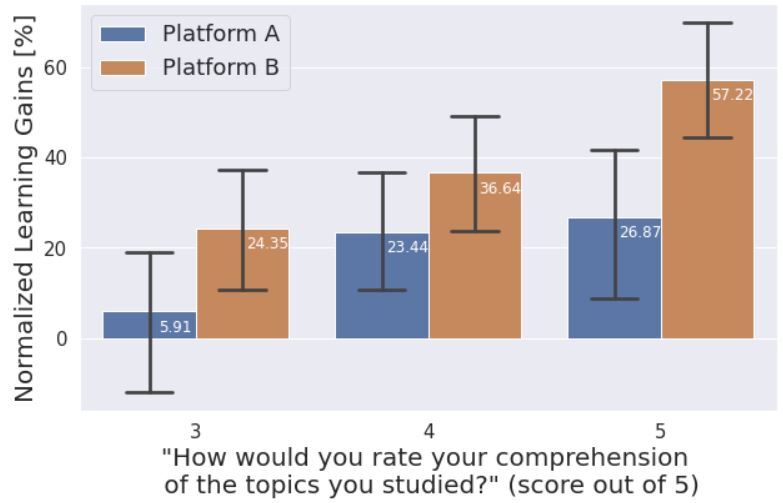}
\caption{Normalized learning gains for each self-assessed comprehension rating with $95\%$ confidence intervals. Only $1$ participant gave a score lower than $3$ (not shown here).} \label{fig:perceived}
\end{figure}

There is a strong correlation between how participants rated their comprehension of the topics studied and their actual learning gains (Q1). This can be seen in Figure \ref{fig:perceived}, which shows the normalized learning gains depending on this subjective rating. The correlation is higher for {\tt Platform B} participants ($r$$=$$0.41$) than for {\tt Platform A} participants ($r$$=$$0.26$). This suggests that {\tt Platform B} gave participants a more accurate understanding of their own knowledge level and helped improve their metacognition. For Q2 and Q3 on perceived learning, no correlation was found with the actual learning gains.

Despite the fact that perceived learning gains are strongly correlated with the actual learning gains for {\tt Platform B}, on average, participants from {\tt Platform B} reported lower levels of comprehension in absolute terms (Q1), and the difference from the {\tt Platform A} average is significant ($p$$<$$0.05$). This directly contradicts the fact that {\tt Platform B} participants obtained higher learning gains on average. In addition, the results for Q2 and Q3 suggest that the {\tt Platform B} participants’ perception of their performance on the post-quiz and their perceived capability to apply the skills they learned are slightly lower than for {\tt Platform A} participants. We note that, in line with previous work, this indicates that students generally struggle to evaluate their own knowledge and skills level~\cite{Brown_13,Brown_15,Crowell}. Furthermore, we hypothesize that the reason for this contradictory result is the presence of some frustrating elements in {\tt Platform B}: for instance, one common source of confusion reported for {\tt Platform B} is the fact that the AI tutor did not always understand participants' answers. On the one hand, the AI tutor on {\tt Platform B} engaging in a dialogue with the participants and providing them with the interactive problem-solving exercises strongly contributes to higher learning gains on {\tt Platform B}. On the other hand, this aspect makes {\tt Platform B} more technically challenging to implement than the more traditional approach taken by {\tt Platform A}, and we hypothesize, in line with previous research~\cite{Lehman,Zhang}, that improvements in the AI tutor's understanding of students' answers and feedback might contribute to higher perceived learning gains in students. Future research should investigate this hypothesis.

\section{Conclusions}
This study compared two popular online learning platforms with respect to the learning outcomes and metacognition induced by the platforms.
The first platform, {\tt Platform A}, is a widely-used learning platform that follows a traditional model for online courses: students on this platform learn by watching lecture videos, reading, and testing their knowledge with multiple choice quizzes.
The second platform, {\tt Platform B}, focuses on active learning and personalization, where each student learns through short lecture videos and interactive problem-solving exercises accompanied by personalized feedback.

We assessed the learning gains of $47$ participants after a $3$-hour long course on linear regression, with $25$ participants taking the course on {\tt Platform A} and $22$ on {\tt Platform B}.
We observed that the average learning gain for {\tt Platform B} is $70.43\%$ higher than on {\tt Platform A} ($p<0.05$).
This supports the hypothesis that participants on {\tt Platform B} have higher learning gains than those who take the course on {\tt Platform A}, because {\tt Platform B} provides a wider and more personalized variety of pedagogical elements to its students.
Furthermore, the average normalized learning gain for {\tt Platform A} participants is $21.73\%$, and $32.43\%$ for {\tt Platform B} participants. With {\tt Platform B} producing $49.24\%$ higher ($p<0.1$) learning gains relative to {\tt Platform A}, the results support the hypothesis that {\tt Platform B} teaches more effectively. However, more data should be collected to validate the same hypothesis w.r.t.\@ normalized learning gains.

In addition, perceived learning gains and metacognition abilities were assessed with a feedback questionnaire filled out by participants after the post-assessment quiz. It was found that {\tt Platform A} participants report higher perceived learning gains despite the fact that their actual learning gains are lower on average. Higher correlation between perceived learning and actual learning gains for {\tt Platform B} participants suggests that {\tt Platform B} induces better metacognition. Lower perceived learning gains on {\tt Platform B} can be explained by the fact that this platform involves more complex teaching elements.
These elements may contribute to higher actual learning gains, they may at the same time also be a source of frustration and confusion for some participants. Future research should look more closely into the factors that affect the perceived learning for participants, and should investigate to what extent the participants benefit from the course being adapted to their learning needs and preferences.


%
%
%
%

\end{document}